\documentclass[twocolumn,prb]{revtex4}
\usepackage[T1]{fontenc}
\usepackage[latin1]{inputenc}
\usepackage{graphicx}
\usepackage{amssymb}

\makeatletter

\usepackage{graphicx}

\makeatother
\begin{document}

\title{$4p$ states and X-Ray Spectroscopy}

\author{Jun-ichi Igarashi}

\affiliation{Faculty of Science, Ibaraki University, Mito, Ibaraki 310-8512, Japan}

\author{Manabu Takahashi}

\affiliation{Faculty of Engineering, Gunma University, Kiryu, Gunma 376-8515,
Japan}

\begin{abstract}
The $4p$ states in transition metals and their compounds usually
play minor roles on their physical quantities. Recent development
of resonant x-ray scattering (RXS) at the $K$-edge of transition
metals, however, casts light on the $4p$ states, because the signals
on orbital and magnetic superlattice spots are brought about by the
modulation in the $4p$ states. The $4p$ states are extending in
solids and thereby sensitive to electronic states at neighboring sites.
This characteristic determines the mechanism of RXS that the intensity
on the orbital superlattice spots are mainly generated by the lattice
distortion and those on magnetic superlattice spots by the coupling
of the $4p$ states with the orbital polarization in the $3d$ states
at neighboring sites. Taking up typical examples for orbital and magnetic
RXS, we demonstrate these mechanisms on the basis of the band structure
calculation. Finally, we study the MCD spectra at the $K$-edge, demonstrating
that the same mechanism as the magnetic RXS is working. 
\end{abstract}
\maketitle

\section{\label{sect.1}Introduction}

X-ray scattering spectroscopy has attracted much attention after high
brilliant synchrotron radiation became available. The scattering intensities
are resonantly enhanced by tuning photon energy close to core excitations.
This resonant x-ray scattering (RXS) is now widely used to investigate
orbital and magnetic orders, since the resonant enhancement makes
it possible to detect the weak intensities on superlattice Bragg spots.

In order to satisfy the Bragg condition on superlattice spots, the
wave length of x-ray has to be an order of lattice spacing. This corresponds
to $K$-edge energies, about $5-10\,\mathrm{keV}$ in transition metals.
Actually the RXS at the $K$-edge has been carried out in transition-metal
compounds for detecting signals on the orbital, magnetic, and charge
superlattice spots.\cite{Caciuffo2002,DiMatteo.S2003,Hill1997,Kubota.M2004,Lorenzo.J.E2005,Mannix2001,Mannix2001b,Martin.H.J2004,Murakami1998,Murakami1998a,Nakamura2004,Nakao.H2000,Namikawa1985,Neubeck1999,Neubeck2001,Noguchi2000,Ohsumi2003,Paolasini1999,Paolasini2002,Stunault1999,Subias.G2004,Zimmermann1999,Zimmermann2001}
The main peak of RXS may be described by a second-order process that
a photon is virtually absorbed by exciting a core electron to unoccupied
\emph{$4p$ states} and then emitted by recombining the excited electron
with the core hole. The problem is that the $4p$ states are not constituting
the orbital order or the magnetic order, and this fact makes the interpretation
of spectra unclear. How is the modulation brought about in the $4p$
states?

\begin{figure}
\includegraphics[%
  scale=0.7]{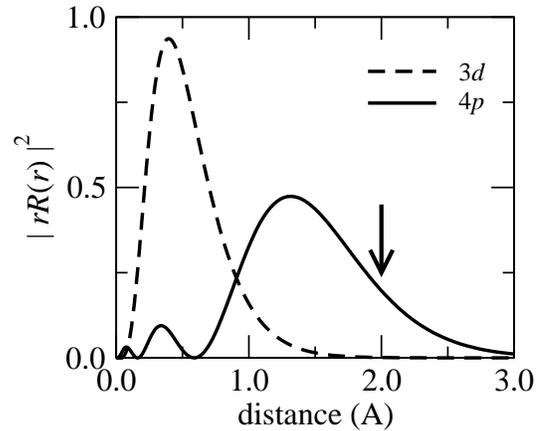}

\caption{\label{fig.4pwave} 4p wave function of an atomic $\mathrm{Mn}^{3+}$
ion with a $1s$ core hole within the Hartree-Fock approximation.
The arrow indicates the distance between the $\mathrm{Mn}$ nucleus
and $\mathrm{O}$ nucleus in $\mathrm{LaMnO}_{3}$. }
\end{figure}

The $4p$ states of transition metals are well extended in solids
and form a broad energy band with their widths $>20\,\mathrm{eV}$
. This fact indicates that simple tight-binding models are not suitable
for the description of the $4p$ bands, but the band structure calculation
is expected to work well. To obtain a feeling of the $4p$ state,
we show in Fig.~\ref{fig.4pwave} the $4p$ wavefunction of a $\mathrm{Mn}^{3+}$
atom within the Hartree-Fock approximation, in comparison with the
$3d$ wavefunction. As seen from the figure, $4p$ electrons have
large probability in the interstitial region, indicating that they
are sensitive to electronic structure at neighboring sites. This observation
leads to important consequences on the mechanism of RXS; the intensity
on orbital superlattice spots are mainly generated by the lattice
distortion changing neighboring atom positions, and those on magnetic
superlattice spots by the coupling of the $4p$ states with the orbital
polarization (OP)%
\footnote{The orbital polarization means the difference in the density of states
with respect to magnetic quantum numbers of the orbital angular momentum,
as a function of energy.%
} in the $3d$ states at neighboring site.

Taking up typical examples for orbital and magnetic RXS, we demonstrate
that those mechanisms are actually working, on the basis of band structure
calculations. Finally we discuss the mechanism of magnetic circular
dichroism (MCD), which proves nearly the same as that of the magnetic
RXS.

The present paper is organized as follows. In Sec. \ref{sect.2},
we study the RXS for both the orbital and magnetic orderings. In Sec.
\ref{sect.3}, we discuss the MCD at the $K$-edge. Section \ref{sect.4}
is devoted to the concluding remarks.

\section{\label{sect.2}Resonant X Ray Scattering}

\begin{figure}
\includegraphics[%
  scale=0.6]{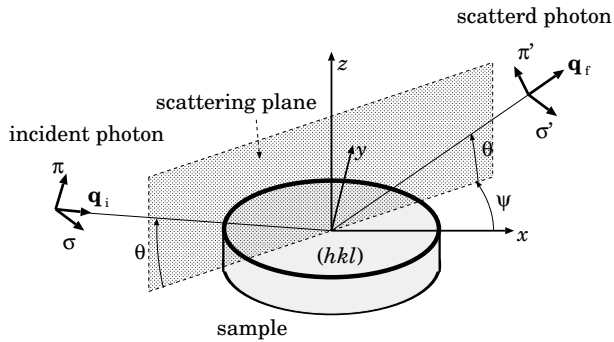}

\caption{\label{fig.geometry} Geometry of x-ray scattering. Incident photon
with wave vector $\mathbf{q}_{\mathrm{i}}$ and polarization $\sigma$
or $\pi$ is scattered into the state with wave vector $\mathbf{q}_{\mathrm{f}}$
and polarization $\sigma^{\prime}$ or $\pi^{\prime}$ at Bragg angle
$\theta$. The sample crystal is rotated by azimuthal angle $\Psi$
around scattering vector $\mathbf{G}=\mathbf{q}_{\mathrm{f}}-\mathbf{q}_{\mathrm{i}}$. }
\end{figure}

We start by a brief summary of the formulas of the RXS intensity.
The scattering tensor can be approximated by a sum of contributions
from each site where a core hole is created, since the $1s$ state
is well localized at transition-metal sites. Then the cross section
for the scattering geometry shown in Fig.~\ref{fig.geometry} is
given by\begin{equation}
\left.\frac{d\sigma}{d\Omega}\right|_{\mu\rightarrow\mu^{\prime}}\propto\left|\sum_{\alpha\alpha^{\prime}}P_{\alpha}^{\prime\mu^{\prime}}M_{\alpha\alpha^{\prime}}\left(\mathbf{G},\omega\right)P_{\alpha^{\prime}}^{\mu}\right|^{2},\end{equation}
 with\begin{eqnarray}
 &  & M_{\alpha\alpha^{\prime}}\left(\mathbf{G},\omega\right)=\frac{1}{\sqrt{N}}\sum_{j}\exp\left(-i\mathbf{G}\cdot\mathbf{r}_{j}\right)\nonumber \\
 &  & \qquad\qquad\;\times\sum_{\Lambda}\frac{\left\langle \psi_{g}\left|x_{\alpha}\left(j\right)\right|\Lambda\right\rangle \left\langle \Lambda\left|x_{\alpha^{\prime}}\left(j\right)\right|\psi_{g}\right\rangle }{\hbar\omega-\left(E_{\Lambda}-E_{g}\right)+i\Gamma}.\label{eq.dipole}\end{eqnarray}
 Here $\mathbf{G}\;\left(=\mathbf{q}_{\mathrm{f}}-\mathbf{q}_{\mathrm{i}}\right)$
is the scattering vector, and $\omega$ is the frequency of photon.
Index $j$ runs over transition-metal sites, the number of which is
denoted as $N$. The cross section would become an order $N$ on superlattice
spots. The dipole operators $x_{\alpha}\left(j\right)$'s are defined
as $x_{1}\left(j\right)=x$, $x_{2}\left(j\right)=y$, and $x_{3}\left(j\right)=z$
in the coordinate frame fixed to the crystal axes with the origin
located at the center of site $j$. The ground-state energy and its
wavefunction are defined as $E_{g}$ and $\left|\psi_{g}\right\rangle $,
respectively. State $\left|\Lambda\right\rangle $ represents the
intermediate state with energy $E_{\Lambda}$. Thus $\left|\Lambda\right\rangle $
has the $p$ symmetry with respect to the origin at the core-hole
site. The $\Gamma$ represents the life-time broadening width of the
core hole, which is usually about $1\,\mathrm{eV}$ in transition-metal
$K$-edges. The $P^{\mu}$ and $P^{\prime\mu^{\prime}}$ are the geometrical
factors for incident and scattered photons with polarization $\mu$
($=\sigma,\pi$) and $\mu^{\prime}$ ($=\sigma^{\prime},\pi^{\prime}$),
respectively.

\subsection{Orbital Order}

The study of orbital order has attracted much interest since the discovery
of colossal magnetoresistance in $\mathrm{La}_{1-x}\mathrm{Sr}_{x}\mathrm{MnO}_{3}$\cite{Urushibara1995}.
The {}``orbital'' is now recognized as an important factor for understanding
the physics of transition-metal compounds. The neutron scattering
has been very useful for the magnetic order, but is not so useful
for the orbital order. Therefore a new method to probe the orbital
order has been awaited. Then, Murakami et al. observed the RXS intensity
at the $K$-edge on orbital superlattice spots, claiming that this
signal is probing the orbital order\cite{Murakami1998,Murakami1998a}.
But it was controversial how the spectra are related to the orbital
order. We discuss this issue in the following.

\subsubsection{$e_{g}$ electron systems}

\begin{figure}
\includegraphics[%
  scale=0.35]{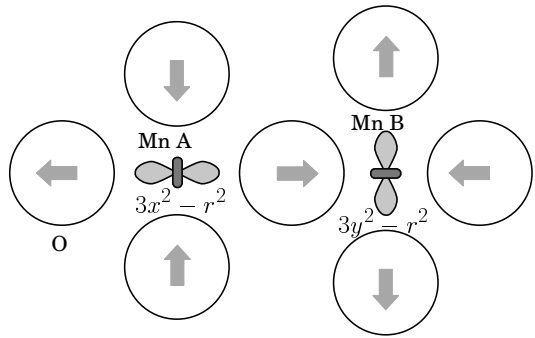}

\caption{\label{fig.ponchi} Schematic view of the orbital order on the $ab$
plane in $\mathrm{LaMnO}_{3}$. Oxygen atoms move along the direction
shown by arrows, which pattern is consistent with the orbital order. }
\end{figure}

We first consider $\mathrm{LaMnO}_{3}$. In a localized picture of
$\mathrm{Mn}^{3+}$ ion, three electrons occupy $t_{2g}$ levels and
one electron occupies $e_{g}$ levels due to the Hund-rule coupling
and the cubic crystal field. The double degeneracy of the $e_{g}$
levels is lifted by the orbital-exchange coupling\cite{Kugel1972}
or the cooperative Jahn-Teller effect.\cite{Kanamori1960} Which is
the driving force does not concern us in this paper. Irrespective
to the driving force, the resulting orbital order and the crystal
distortion are those shown in Fig.~\ref{fig.ponchi}; the $3x^{2}-r^{2}$
and $3y^{2}-r^{2}$ orbitals are alternately arranged on the $ab$
plane with oxygens positions shifted toward directions indicated by
arrows.

\begin{figure}
\includegraphics[%
  scale=0.8]{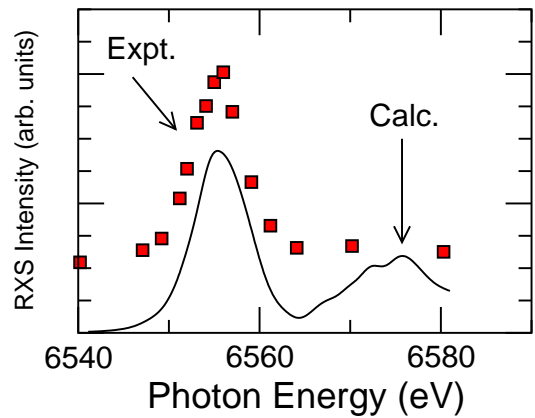}

\caption{\label{fig.lamn} Calculated RXS intensity on an superlattice spot
at the $K$-edge in $\mathrm{LaMnO}_{3}$, in comparison with the
experiment. }
\end{figure}

As easily inferred from Eq.~(\ref{eq.dipole}), the intermediate
states ($4p$ bands) have to be modulated in accordance with the orbital
order for the RXS intensity not to vanish on the superlattice spots.
If the anisotropic terms of the $3d-4p$ Coulomb interaction is effective,
the $4p_{x}$ level becomes higher than the $4p_{y}$ and $4p_{z}$
levels for $\mathrm{Mn}$ sites with the $3x^{2}-r^{2}$ state, while
the $4p_{y}$ level becomes higher than the $4p_{x}$ and $4p_{z}$
levels for $\mathrm{Mn}$ sites with $3y^{2}-r^{2}$ state. On the
other hand, if the hybridization between the $4p$ states and oxygen
$2p$ states at neighboring sites is effective, the shifts of oxygen
positions make the $4p_{y}$ level higher than $4p_{x}$ and $4p_{z}$
levels for $\mathrm{Mn}$ sites with the $3x^{2}-r^{2}$ state while
the $4p_{x}$ level higher than $4p_{x}$ and $4p_{z}$ levels for
$\mathrm{Mn}$ sites with $3y^{2}-r^{2}$ state. The two origins lead
to the level shift opposite to each other. In solids, the $4p$ states
are not those localized levels but form energy bands with width of
order $20\,\mathrm{eV}$. Nevertheless, the tendency mentioned above
on the $4p$ levels would still hold in solids. At the early stage,
Ishihara et al. interpreted the RXS signal on the orbital superlattice
spots in $\mathrm{LaMnO}_{3}$ by the {}``Coulomb'' mechanism.\cite{Ishi1998a,Ishi1998b}
By this mechanism, we could say that the modulation in $4p$ bands
is directly caused by the $3d$ orbital order. However, subsequent
studies based on band structure calculations have revealed that the
Jahn-Teller distortion (JTD) gives rise to large RXS intensities.\cite{Benfatto1999,Elfimov1999,Taka1999}
We show in Fig.~\ref{fig.lamn} the calculated spectra of the RXS
intensity on an orbital superlattice spot in $\mathrm{LaMnO}_{3}$.\cite{Taka1999}
The calculation is based on the local density approximation (LDA)
within the muffin-tin (MT) approximation on the lattice parameters
determined from the experiment. This approximate scheme leads to a
small energy gap ($0.2\,\mathrm{eV}$) with a small orbital polarization.
The {}``orbital\char`\"{} in this subsection means the levels with
real wavefunctions such as $3x^{2}-r^{2}$, $3y^{2}-r^{2}$, $p_{x}$,
$p_{y}$, $p_{z}$. In spite of this shortcoming for the $3d$ bands,
we expect that the $4p$ bands are well described in the present calculation,
since they have energies $\sim15\,\mathrm{eV}$ higher than the $3d$
bands and thereby the details of the $3d$ bands are irrelevant. Actually,
it was shown that the LDA$+U$ method, which predicts a sufficient
energy gap with large orbital polarization, gives the $4p$ bands
and the RXS spectra nearly the same as those given by the LDA method.\cite{Benedetti2001}
Since the MT approximation averages the potential coming from orbitally
polarized $3d$ states, it works to eliminate the effect of the anisotropic
terms of the $3d$-$4p$ Coulomb interaction on the $4p$ states.
Thus, the calculated intensity is regarded purely generated by the
JTD. The effect of the Coulomb interaction has been examined and estimated
about two order of magnitude smaller than the JTD effect.\cite{Taka1999,Benedetti2001}
This finding is quite reasonable, since the $4p$ states are well
extended and sensitive to electronic states at neighboring sites.

Another typical example is $\mathrm{KCuF}_{3}$. Each $\mathrm{Cu}^{2+}$
ion takes the $3d^{9}$-configuration, in which one hole occupies
one of the $e_{g}$ states. By the same reason as in $\mathrm{LaMnO}_{3}$,
the orbital order and the JTD take place. The band structure calculations
\cite{Taka2003a,Binggeli.N2004} have been applied to evaluate the
RXS intensity on the orbital superlattice spots in good agreement
with the experiment,\cite{Caciuffo2002} clarifying that it arises
mainly from the JTD.

As mentioned above, the $4p$-level shift by the JTD mechanism would
be opposite to that by the Coulomb mechanism. Unfortunately, such
difference in shift cannot be distinguished in antiferro-type orbital
orders, due to a cancellation between different sublattices. The situation
may be different in ferro-type orbital orders, but the actual detect
of the difference is quite difficult because of the overlapping fundamental-reflection
intensities. To circumvent this difficulty, Kiyama et al.,\cite{Kiyama.T2003}
Ohsumi et al.\cite{Ohsumi2003} and Nakamura et al.\cite{Nakamura2004}
devised an interference technique and applied it to $\mathrm{La}_{0.45}\mathrm{Sr}_{0.55}\mathrm{MnO}_{3}$
and $\mathrm{La}_{0.60}\mathrm{Sr}_{0.40}\mathrm{MnO}_{3}$ multilayers.
They have found that the level shift is indeed consistent with the
JTD mechanism and opposite to the shift predicted by the Coulomb mechanism.

\subsubsection{$t_{2g}$ electron systems}

In $t_{2g}$ electron systems such as $\mathrm{YTiO}_{3}$ and $\mathrm{YVO}_{3}$
where $t_{2g}$ orbitals are partially occupied, the JTD is usually
smaller than $e_{g}$ electron systems such as $\mathrm{LaMnO}_{3}$
and $\mathrm{KCuF}_{3}$. In such a situation, one may think the effect
of Coulomb interaction important. This is not the case, though. Taking
up the RXS in $\mathrm{YVO}_{3}$, we demonstrate that the RXS spectra
are mainly determined by the lattice distortion.\cite{Taka2002}

This material takes three phases\cite{Kawano.H1994}: (a) a G-type
antiferromagnetic state is developed with a moderate size of the JTD
for $T<77$ K (the low-temperature (LT) phase); (b) a C-type antiferromagnetic
state is developed with a small JTD for $77$ K $<T<118$ K (the intermediate-temperature
(IT) phase); (c) no magnetic order is developed but with a small JTD
for $118$ K $<T$ (the high-temperature (HT) phase). In the three
phases, the $\mathrm{GdFeO}_{3}$-type distortion, the rotation and
tilt of the $\mathrm{VO}_{6}$ octahedron, is intrinsically present
in addition to the JTD, which makes the crystal belong to the space
group $Pbnm$.

\begin{figure}
\includegraphics[%
  width=0.45\textwidth]{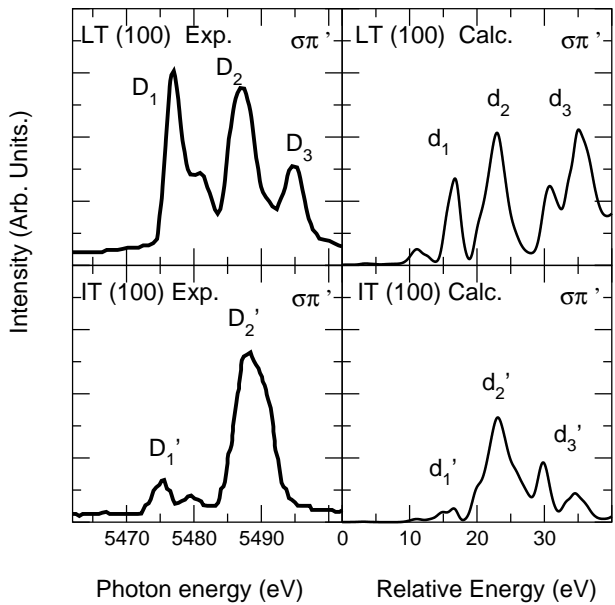}

\caption{\label{fig.yvo3} RXS spectra on the $(100)$ reflection at the $K$-edge
in $\mathrm{YVO}_{3}$. Right and left panels show the calculated
spectra and the experimental data from Ref.\onlinecite{Noguchi2000}
($\Psi=0$), respectively. Upper and lower panels are for the LT and
IT phases, respectively. }
\end{figure}

Noguchi et al.\cite{Noguchi2000} have observed the RXS spectra at
the $K$-edge of $\mathrm{V}$, which are shown on the left panels
in Fig.~\ref{fig.yvo3}; three peaks are seen as a function of photon
energy in the LT phase, while two peaks are suppressed with surviving
one peak labeled {}``$\mathrm{D}_{2}$'' in the IT phase. The RXS
spectra are calculated based on the LDA and the MT approximation with
the lattice parameters determined by the experiment. The result on
the $(100)$ reflection is shown on the right panels in Fig.~\ref{fig.yvo3}.\cite{Taka2002}
Three peak structure is well reproduced. The spectra change drastically
in the IT phase; peaks $\mathrm{d}_{1}$ and $\mathrm{d}_{3}$ are
now strongly suppressed, although peak $\mathrm{d}_{2}$ keeps its
strength. This behavior is in good agreement with the experiment.
In addition, we have calculated the RXS spectra in an \emph{imaginary}
cubic crystal structure which has no $\mathrm{GdFeO}_{3}$-type distortion
but the JTD of the same type as in the real crystal, and have found
that peak $\mathrm{d}_{2}$ is strongly suppressed but peaks $\mathrm{d}_{1}$
and $\mathrm{d}_{3}$ keep their strength.\cite{Taka2002} This indicates
that the contribution of the JTD is mainly concentrated to peaks $\mathrm{D}_{1}$
and $\mathrm{D}_{3}$ in the (100) reflection, while that of the $\mathrm{GdFeO}_{3}$-type
distortion is distributed to other peaks particularly peak $\mathrm{D}_{2}^{\prime}$.
Calculating the RXS spectra with changing magnetic structures but
with the same lattice distortion, we have confirmed that the magnetic
structure affects little on the RXS spectra. This is consistent with
the experimental finding that the RXS spectra in the HT phase are
quite similar to those in the IT phase, because the main difference
between the IT and HT phases are the presence or absence of magnetic
order.

The experiment was first interpreted by the Coulomb mechanism.\cite{Ishihara.S2002}
This mechanism is unable to explain the three-peak structure and its
difference between the LT and IT phases. The band calculation explains
naturally most aspects of the spectra, Thereby we may conclude that
the $4p$ states are sensitive enough to the lattice distortion and
that the RXS spectra are mainly controlled by the lattice distortion
even in $t_{2g}$ systems.

\subsection{Magnetic Order}

In magnetic materials, RXS signals are observed also on magnetic superlattice
spots. The spin polarization alone could not give rise to the RXS
intensity, because both the majority and minority spin states contribute
to the scattering amplitude and thereby the scattering amplitudes
are the same on all sites. The orbital polarization (OP) with respect
to symmetries $p_{+}=\frac{-1}{\sqrt{2}}\left(p_{x}+ip_{y}\right)$
and $p_{-}=\frac{1}{\sqrt{2}}\left(p_{x}-ip_{y}\right)$ is necessary
not to vanish the intensity. This is achieved by the spin-orbit interaction
(SOI), which makes the orbital degrees of freedom couple to the spin
polarization. The scattering amplitude $M_{\alpha,\alpha'}$ has an
antisymmetric form. Now a question is that how the spin order in the
$3d$ bands influence the OP in the $4p$ bands. A natural answer
lies again on the extended nature of the $4p$ states. The OP is induced
by the coupling to the OP in the $3d$ bands at neighboring sites
through the $p$-$d$ hybridization. We explain this scenario by taking
up the spin density wave (SDW) phase in the $\mathrm{Cr}$ metal.
\cite{Taka2004Cr}

\begin{figure}
\includegraphics[%
  scale=0.45]{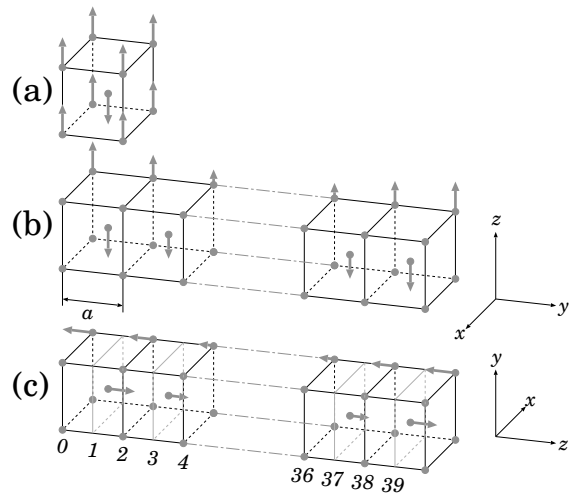}

\caption{\label{fig.structure} Schematic view of the SDW state with $Q_{\mathrm{SDW}}=\frac{2\pi}{a}\frac{19}{20}$
with $a$ the lattice constant of simple bcc structure. (a) A simple
antiferromagnetic state. (b) A TSDW state. (c) An LSDW state. Integers
indicate Cr site numbers. }
\end{figure}

The $\mathrm{Cr}$ metal forms a bcc lattice structure.\cite{Fawcett1988}
As cooling its magnetic state turns into a transverse spin density
wave (TSDW) state at N\'{e}el temperature $T_{\mathrm{N}}=311\,\mathrm{K}$
and into a longitudinal spin density wave (LSDW) state at spin-flip
temperature $T_{\mathrm{SF}}=122\,\mathrm{K}$. In the TSDW (LSDW)
state, magnetic moments are perpendicular (parallel) to the SDW propagation
vector, as schematically shown in Fig.~\ref{fig.structure}.

Although the wavelength of the SDW is incommensurate with the lattice
periodicity, we approximate the SDW wavelength to be $\lambda_{\mathrm{SDW}}=20a$
with $a$ the bcc lattice constant. Thereby we are dealing with a
large unit cell containing 40 inequivalent $\mathrm{Cr}$ atoms. This
period is very close to the observed value at $T_{\mathrm{SF}}$.
We carry out the band calculation self-consistently with the large
unit cell.\cite{Taka2004Cr} Within the LDA method and the MT approximation,
we obtain the LSDW and TSDW phases as stable solutions with the magnetic
moment written as \begin{eqnarray}
\mu_{j} & = & M_{1}\cos\left(\mathbf{Q}_{\mathrm{SDW}}\cdot\mbox{\boldmath$\tau$}_{j}\right)+M_{3}\cos\left(3\mathbf{Q}_{\mathrm{SDW}}\cdot\mbox{\boldmath$\tau$}_{j}\right)\nonumber \\
 &  & +M_{5}\cos\left(5\mathbf{Q}_{\mathrm{SDW}}\cdot\mbox{\boldmath$\tau$}_{j}\right)+\cdots.\label{eq:moment_fourie}\end{eqnarray}
 Here \textbf{$\mathbf{Q}_{\mathrm{SDW}}$} is a propagation vector
of the SDW with $\mbox{\boldmath$\tau$}_{j}$ denoting the position
vector of the $j$th site ($j=0,1,\cdots,39$ in Fig.~\ref{fig.structure}).
The magnetic moment is made up of the $3d$ spin moment; the amplitudes
in Eq.~(\ref{eq:moment}) are evaluated as $M_{1}^{3d}=0.393\hbar$,
$M_{3}^{3d}=-0.026\hbar$, $M_{5}^{3d}=0.0025\hbar$ in the LSDW state.
These values are consistent with the reported band calculations \cite{Hirai1998,Hafner2002}
and the experiment.\cite{Mannix2001}

\begin{figure}
\includegraphics[%
  scale=0.7]{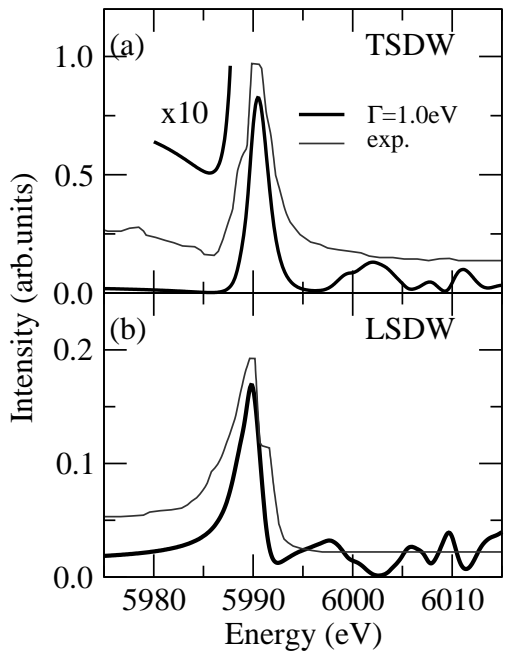}

\caption{\label{fig.cr1} (a) Calculated magnetic scattering intensities for
$\sigma\pi^{\prime}$ channel on the SDW Bragg spot $\mathbf{Q}_{\mathrm{TSDW}}$
in the TSDW state. (b) Those on the spot $\mathbf{Q}_{\mathrm{LSDW}}$
in the LSDW state. The energy of the $1s$ core level is assumed to
be $\epsilon_{1s}=\epsilon_{\mathrm{F}}-5988.4\,\mathrm{eV}$ with
$\Gamma=1.0\,\mathrm{eV}$. The experimental curve is traced from
Ref.\onlinecite{Mannix2001}. }
\end{figure}

Then, adding the SOI term $\frac{1}{r}\frac{d}{dr}V\left(r\right)\ell_{z}s_{z}$
to the MT potential $V(r)$, we calculate the eigenvalues and eigenfunctions,
and finally calculate the magnetic RXS spectra. Figure \ref{fig.cr1}
shows the calculated results in the LSDW and TSDW states, \cite{Taka2004Cr}
in comparison with the experiment\cite{Mannix2001}. The calculation
reproduces well not only the resonant behavior but also the Fano-type
dip around $\hbar\omega=5986\,\mathrm{eV}$ in the TSDW state. (The
Fano phenomena arise from the interference between the resonant and
non-resonant processes; we have omitted the description of the non-resonant
term.)

\begin{figure}
\includegraphics[%
  scale=0.7]{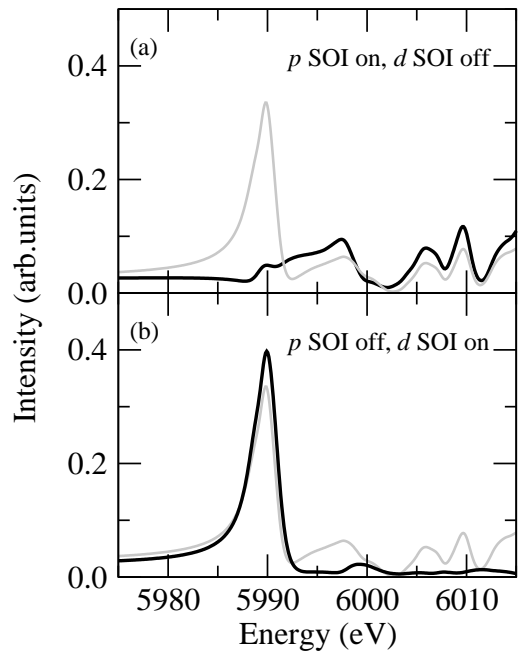}

\caption{\label{fig.cr2} Scattering intensity on the SDW spot $\mathbf{Q}_{\mathrm{LSDW}}$
in LSDW state. (a) The SOI in $p$ state is turned on and the SOI
in $d$ state is turned off. (b) The SOI in $p$ state is turned off
and the SOI in $d$ state in turned on. The gray lines represent the
intensities calculated with both of the $p$ SOI and $d$ SOI being
turned on. }
\end{figure}

How is the OP induced in the $4p$ bands? This question is answered
by calculating the RXS spectra with turning on and off the SOI on
the $p$ and $d$ states selectively. The result is shown in Fig.~\ref{fig.cr2}.\cite{Taka2004Cr}
When the SOI is turned off only on the $d$ states, the resonant main
peak at $\hbar\omega=5990\,\mathrm{eV}$ disappears, while the resonant
structures at higher energies change little. On the other hand, when
the SOI is turned off only on the $p$ states, the main peak dose
not change, while the resonant structure at higher energies almost
disappears. Note that the $p$ and $d$ states do not hybridize each
other at the same site because the SDW wavelength is so long that
each site almost keeps the inversion symmetry. Therefore, the above
finding indicates that the OP in the $4p$ bands near the Fermi level,
which gives rise to the main RXS peak, is mainly induced by the OP
in the $3d$ states at neighboring sites through the $p$-$d$ hybridization.
Since the $3d$ density of states is concentrated near the Fermi level,
the $p$-$d$ hybridization may work effectively. It should be noted
here that the orbital moment, which is given by the sum of the OP
in the occupied levels, is found very small, but that the OP in the
$3d$ bands as a function of energy is found large (see Fig.~8 in
Ref.\onlinecite{Taka2004Cr}).

The fact that the OP in the $4p$ bands is induced by the hybridization
with the states at neighboring sites may be more clearly seen in the
RXS for $\mathrm{UGa}_{3}$, where a large peak at the $K$-edge of
$\mathrm{Ga}$ is observed on the antiferromagnetic superlattice spots.\cite{Mannix2001b}
The OP in the $4p$ states of $\mathrm{Ga}$ is induced through the
hybridization with the $5f$ states of U. Since the OP in the $5f$
states of U ls large, the induced OP in the $4p$ states of $\mathrm{Ga}$
is also large, and this gives rise to the large peak. The band structure
calculation has confirmed this scenario\cite{Usuda2004} ; when the
SOI is turned off only on the 5f states, the peak is strongly suppressed.

In transition-metal compounds, the $4p$ bands are usually located
more than $10\,\mathrm{eV}$ above the $3d$ bands. A typical example
is $\mathrm{NiO}$.\cite{Hill1997,Neubeck2001} In such a situation,
the effect of the hybridization with the $3d$ bands is not large.
The band calculation has revealed that the OP in the $4p$ bands is
mainly induced by the SOI on the $4p$ states.\cite{Usuda2004b} This
does not simply mean that the coupling to neighboring $3d$ states
is unimportant, because the spin moment in the $4p$ bands is induced
by the coupling to the spin moment in the $3d$ band. At any rate,
the inducing mechanism is not simple here. As pointed out in the magnetic
RXS, the same situation takes place in high energy states of the $4p$
bands in $\mathrm{Cr}$ metal.

\section{\label{sect.3}Magnetic Circular Dichroism}

We first summarize the formulas of MCD spectra. By neglecting the
core-hole potential, the absorption coefficient for the right-handed
$(+)$ and left-handed $(-)$ circular polarizations are calculated
from \begin{eqnarray}
\mu_{\pm}(\omega) & \propto & \sum_{n,\mathbf{k}}\left|\int r^{2}dr\int\mathrm{d}\Omega\phi_{n,\mathbf{k}}\left(\mathbf{r}\right)^{*}rY_{1,\pm1}(\Omega)R_{1s}(r)\right|^{2}\nonumber \\
 & \times & \theta(\epsilon_{n,\mathbf{k}}-\epsilon_{\mathrm{F}})\frac{\Gamma}{\pi}\frac{1}{\left(\hbar\omega-\epsilon_{n,\mathbf{k}}\right)^{2}+\Gamma^{2}},\label{eq.xmcd1}\end{eqnarray}
 where the spherical harmonic function $Y_{1,\pm1}\left(\Omega\right)$
is defined with the quantization axis along the propagating direction
of photon. The step function $\theta\left(x\right)$ ensures that
the sum is taken over states above the Fermi level. The $R_{1s}$
represents the $1s$ wave function of transition metals, and $\phi_{n,\mathbf{k}}$
represents the wave function with the band index $n$, wave-vector
$\mathbf{k}$ and energy $\epsilon_{n,{\textbf{k}}}$. Equation (\ref{eq.xmcd1})
indicates clearly that only the $p$ symmetric states contribute the
intensity. Since the $4p$ states are extending, the assumption of
neglecting the core-hole potential may be justified. The total absorption
coefficient and the MCD spectra are defined by \begin{eqnarray}
\mu_{0}\left(\omega\right) & = & \left[\mu_{+}\left(\omega\right)+\mu_{-}\left(\omega\right)\right]/2\\
\mu_{c}\left(\omega\right) & = & \left[\mu_{+}\left(\omega\right)-\mu_{-}\left(\omega\right)\right].\label{eq.xmcd2}\end{eqnarray}

As is clear from Eqs.~(\ref{eq.xmcd1}) and (\ref{eq.xmcd2}), the
MCD spectra at the $K$-edge probes the OP in the $4p$ bands, which
situation is the same as the magnetic RXS at the $K$-edge. The main
difference is that RXS could provide information on spatially modulated
magnetic order such as antiferromagnetic states, while MCD is restricted
on ferromagnetic states. In any case, the mechanism of inducing the
OP in the $4p$ bands is closely related.

Ebert et al.\cite{Ebert1988} made an LDA calculation of the MCD spectra
for Fe in good agreement with the experiment.\cite{Schutz1987} A
subsequent calculation for $\mathrm{Ni}$,\cite{Stahler1993} however,
gave merely semi-quantitative agreement with the experiment near the
photothreshold. In addition, the underlying mechanism for the MCD
was not clarified. One of the present author and Hirai\cite{Iga1994a,Iga1996b}
made a tight-binding calculation of the MCD spectra for $\mathrm{Fe}$
and $\mathrm{Ni}$, where parameters were determined so as to fit
the LDA band structure calculation. They not only reproduced well
Ebert et al's result for Fe, but also obtained the spectra for $\mathrm{Ni}$
in good agreement with the experiment. They clarified the underlying
mechanism by switching on and off the SOI selectively on the $4p$
and $3d$ states; the MCD spectra are mainly generated by the OP in
the $3d$ states at neighboring sites through the $p$-$d$ hybridization.
This mechanism has recently been reconfirmed by the calculation of
the MCD spectra in $\mathrm{Mn}_{3}\mathrm{GaC}$ on the basis of
the band calculation. \cite{Taka2003b} This example is discussed
in detail in the following.

The $\mathrm{Mn}_{3}\mathrm{GaC}$ takes an {}``inverse'' perovskite
structure.\cite{Fruchart.D1978} The magnetic property is that the
antiferromagnetic phase in low temperatures turns into a ferromagnetic
phase in high temperatures through a first-order transition at $168\,\mathrm{K}$.\cite{Uemoto.S2001}
We carry out the band calculation assuming a ferromagnetic phase,\cite{Taka2003b}
although an antiferromagnetic phase may be stabler than the ferromagnetic
phase. The magnetization is obtained as $\sim0.8\mu_{\mathrm{B}}$
per $\mathrm{Mn}$. The SOI is introduced in the same way as in the
calculation of the magnetic RXS spectra.

\begin{figure}
\includegraphics[%
  scale=0.8]{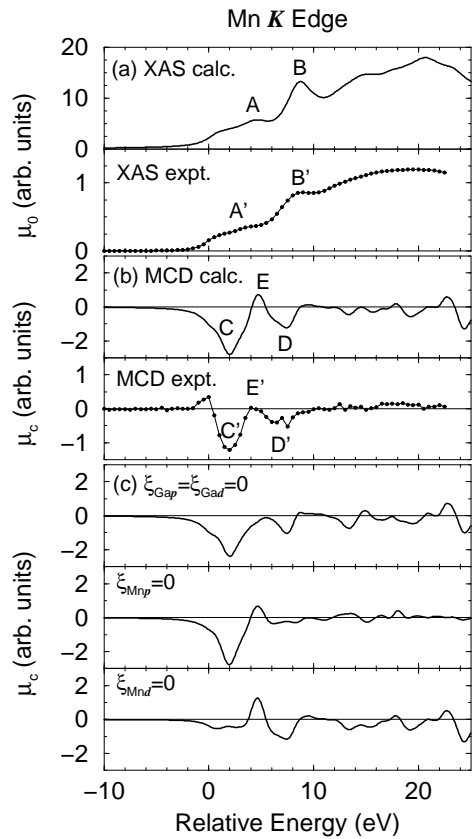}

\caption{\label{fig.xmcd1} Total absorption coefficient $\mu_{0}(\omega)$
and MCD spectra $\mu_{c}(\omega)$ an the $K$-edge of $\mathrm{Mn}$,
as a function of photon energy. The origin of energy corresponds to
the excitation to the Fermi level. The experimental data are for a
powdered sample in the ferromagnetic phase at $T=200$ K (Ref.\onlinecite{Taka2003b}).
(a) $\mu_{0}(\omega)$. (b) $\mu_{c}(\omega)$. (c) $\mu_{c}(\omega)$
calculated with tuning off the SOI on all the states at Ga sites,
on the $p$ symmetric states at $\mathrm{Mn}$ sites, and on the $d$
symmetric at $\mathrm{Mn}$ sites, respectively. ($\xi_{Ab}=0$ on
panels indicates that the SOI on the $b$ symmetric states at A sites
is turned off.) }
\end{figure}

Figure \ref{fig.xmcd1} shows the calculated MCD spectra at the $K$-edge
of $\mathrm{Mn}$, in comparison with the experiment.\cite{Uemoto.S2001}
The calculated value of $\mu_{c}(\omega)$ is divided by the value
of $\mu_{0}(\omega)$ at $\hbar\omega=\epsilon_{\mathrm{F}}+20\,\mathrm{eV}$,
while the experimental MCD spectra are divided by the value of the
total absorption coefficient at the energy about $40\,\mathrm{eV}$
higher than the threshold. Both $\mu_{0}(\omega)$ and $\mu_{c}(\omega)$
are in good agreement with the experiment. For the total absorption
coefficient, peaks A and B correspond well with shoulders $\mathrm{A}^{\prime}$
and $\mathrm{B}^{\prime}$ in the experimental curve. For the MCD
spectra, dips $\mathrm{C}$, $\mathrm{D}$, and peak $\mathrm{E}$
correspond well with the experimental ones $\mathrm{C}^{\prime}$,
$\mathrm{D}^{\prime}$, and $\mathrm{E}^{\prime}$. The underlying
mechanism is clarified by calculating the MCD spectra at $\mathrm{Mn}$
sites with turning off the SOI on several specified states. The top
panel among three panels of Fig.~\ref{fig.xmcd1}(c) shows the MCD
spectra with turning off the SOI on all the states at $\mathrm{Ga}$
sites. The spectra remain similar except for a slight suppression
of peak E, indicating that the OP at $\mathrm{Ga}$ sites have little
influence on the MCD spectra at $\mathrm{Mn}$ sites. The middle panel
of Fig.~\ref{fig.xmcd1}(c) shows the MCD spectra with turning off
the SOI only on the $p$ symmetric states at $\mathrm{Mn}$ sites.
Dip $\mathrm{C}$ keeps the similar shape, while dip $\mathrm{D}$
almost vanishes. This indicates that the OP corresponding to dip $\mathrm{D}$
is induced by the spin polarization in the $p$ symmetric states through
the SOI in the $4p$ states. The bottom panel shows the MCD spectra
with turning off the SOI only on the $d$ symmetric states at $\mathrm{Mn}$
sites. The intensity of dip $\mathrm{C}$ is drastically reduced,
but dip $\mathrm{D}$ and peak $\mathrm{E}$ remains similar, indicating
that the OP in the $3d$ bands gives rise to dip $\mathrm{C}$. Within
the MT approximation, the $3d$ OP cannot polarize the $p$ orbital
in the same $\mathrm{Mn}$ site, because the $p$-$d$ Coulomb interaction
is spherically averaged inside the MT sphere. Thus we conclude that
the OP corresponding to dip $\mathrm{C}$ is induced by the $3d$
OP at \emph{neighboring} $\mathrm{Mn}$ sites through the $p$-$d$
hybridization. Although the figure is not shown (see Fig.~3 in Ref.\onlinecite{Taka2003b}),
the MCD spectra are also observed at the $K$-edge of $\mathrm{Ga}$.
The same analysis has been applied to this case, clarifying that the
OP in the $4p$ states of $\mathrm{Ga}$ is mainly induced by the
OP in the $3d$ states of $\mathrm{Mn}$ at neighboring sites through
the $p$-$d$ hybridization.

\section{\label{sect.4} Concluding remarks}

We have studied the mechanisms of RXS and MCD at the $K$-edge in
transition metals and their compounds. The band structure calculations
have proved powerful enough to reproduce well the experimental spectra.

Analyzing in detail the cases for $\mathrm{LaMnO}_{3}$ and $\mathrm{YVO}_{3}$,
we have demonstrated that the RXS spectra on orbital superlattice
spots arise mainly from the lattice distortion. Since the extended
nature of the $4p$ states naturally leads to sensitiveness to electronic
structures at neighboring sites, this conclusion looks quite natural.
The {}``Coulomb mechanism''\"{} is found irrelevant to the orbital
RXS. In materials with extremely small lattice distortions, however,
the anisotropic term of the intraatomic Coulomb interaction may play
an important role. RXS spectra have been observed on quadrupole ordering
spots at the $\mathrm{Ce}$ $L_{3}$-edge in $\mathrm{CeB}_{6}$,\cite{Nakao.H2001}
in which any sizable lattice distortions have not been observed. A
detailed analysis\cite{Nagao.T2001,Igarashi.J2002} has shown that
the spectra are generated by modulating $5d$ states through the intraatomic
Coulomb interaction between the $5d$ states and the orbitally polarized
$4f$ states.

In addition, analyzing the SDW phase of $\mathrm{Cr}$, we have demonstrated
that the RXS spectra on magnetic superlattice spots arise from the
OP in the $4p$ states, which is induced by the OP in the $3d$ states
at neighboring sites through the $p$-$d$ hybridization.

Finally we have discussed the mechanism of MCD spectra by taking the
example of $\mathrm{Mn}_{3}\mathrm{GaC}$. We have emphasized a close
relation to the mechanism of the MCD spectra at the $K$-edge. Again,
it has been stressed that the $4p$ states are extended and sensitive
to electronic structures at neighboring sites.

We have not discussed the pre-edge peak at about $10\,\mathrm{eV}$
below the main peak. The intensity is usually one or two orders of
magnitude smaller than the main-peak intensity. In the electric dipole
transition, the relevant states must have the $p$ symmetry with respect
to the core-hole site. Since the energy of the peak corresponds to
the excitation to the $3d$ states, the $p$ symmetric states have
to be constructed by the $3d$ states at neighboring sites. On the
other hand, in the electric quadrupole transition, the relevant states
have the $d$ symmetry with respect to the core-hole sites. Therefore
the $d$ symmetric states are the $3d$ states themselves. We could
distinguish two processes by their azimuthal angle dependences of
the spectra. The pre-edge peak observed on orbital superlattice spots
in $\mathrm{YTiO}_{3}$ shows the same azimuthal angel dependence
as the main peak, and thereby it is concluded to come from the dipole
transition.\cite{Kubota.M2004} A possible pre-edge peak on the orbital
superlattice spots in $\mathrm{LaMnO}_{3}$ has been analyzed in detail
by the band calculation,\cite{Taka2000} although the experimental
confirmation has not been done yet. It is concluded that the dipole
transition dominates the intensity. On the other hand, it is found
that the pre-edge spectra on the magnetic superlattice spots in $\mathrm{NiO}$
arises from the quadrupole transition. The analysis based on the band
calculation\cite{Usuda2004b} has reproduced well the spectral pattern
arising from the interference between the non-resonant process and
the quadrupole process. Note that the distinction between $p$ and
$d$ symmetries would lose the meaning in crystal structures without
inversion symmetry, such as $\mathrm{V}_{2}\mathrm{O}_{3}$ and $\mathrm{Cr}_{2}\mathrm{O}_{3}$.\cite{Fabrizio.M1998,Paolasini1999,Tanaka.A2002,Elfimov2002,Lovesey.S.W2002,Joly.Y2004}
An interference between the dipole and quadrupole processes may become
important.

If the RXS is available at the $L_{2,3}$-edge in transition metals,
the spectra may be a direct probe to orbital and magnetic orders,\cite{Castleton.C.W.M2000,Wilkins2003,Wilkins2003b}
because the electric dipole transition excites an electron from the
$2p$ core to the $3d$ states. The corresponding x-ray wave length
is much larger than the lattice spacing, though. Recently, such attempts
have been in progress to study orders with large periods and nanostructures.

Finally we comment the RXS at the $M_{4,5}$-edge in actinide compounds.\cite{Isaacs.E.D.1990,Paixao.J.A2002,Lovesey.S.W.2003,Nagao.T2005}
The dipole transition excites an electron from the $3d$ core to the
$5f$ states, and the corresponding x-ray wavelength is an order of
lattice spacing. Therefore the RXS may become a direct probe to the
$5f$ states in order to study multipole orders such as quadrupole
and octupole orders.

\begin{acknowledgments}
We would like to thank N. Hamada, K. Hirai, T. Nagao, and M. Usuda
for valuable discussions. This work was partially supported by a Grant-in-Aid
for Scientific Research from the Ministry of Education, Culture, Sports,
Science, and Technology, Japan. 
\end{acknowledgments}
\bibliography{Bibfile}

\end{document}